\documentclass[aps,twocolumn,prl,showpacs,superscriptaddress]{revtex4-1}

\usepackage{graphicx}
\usepackage{dcolumn}
\usepackage{bm}
\usepackage[T1]{fontenc}
\usepackage[latin9]{inputenc}
\usepackage{textcomp}
\usepackage{amsmath}
\usepackage{amssymb}
\usepackage{units}
\usepackage{xcolor}
\usepackage[normalem]{ulem}
\usepackage{siunitx}
\usepackage{datetime}

\begin{document}

\title{Thermal decoherence of a nonequilibrium polariton fluid}

\author{Sebastian Klembt}
\affiliation{Univ. Grenoble Alpes, CNRS, Institut N\'{e}el, 38000 Grenoble, France}
\altaffiliation[Present address: ]{Technische Physik, Universit\"at W\"urzburg, Am Hubland, D-97074 W\"urzburg, Germany}
\author{Petr Stepanov}
\affiliation{Univ. Grenoble Alpes, CNRS, Institut N\'{e}el, 38000 Grenoble, France}
\author{Thorsten Klein}
\affiliation{University of Bremen, P.O. Box 330440, 28334 Bremen, Germany}
\altaffiliation[Present address: ]{BIAS, Bremen Institute of Applied Beam Technology GmbH, Klagenfurter Str. 5, D-28359 Bremen, Germany}
\author{Anna Minguzzi}
\affiliation{Univ. Grenoble Alpes, CNRS, LPMMC, 38000 Grenoble, France}
\author{Maxime Richard}
\affiliation{Univ. Grenoble Alpes, CNRS, Institut N\'{e}el, 38000 Grenoble, France}

\pacs{71.36.+c,05.70.Ln,67.10.Jn}


\begin{abstract}
Exciton-polaritons constitute a unique realization of a quantum fluid interacting with its environment. Using Selenide based microcavities, we exploit this feature to warm up a polariton condensate in a controlled way and monitor its spatial coherence. We determine directly the amount of heat picked up by the condensate by measuring the phonon-polariton scattering rate and comparing it with the loss rate. We find that upon increasing the heating rate, the spatial coherence length decreases markedly, while localized phase structures vanish, in good agreement with a stochastic mean field theory. From the thermodynamical point-of-view, this regime is unique as it involves a nonequilibrium quantum fluid with no well-defined temperature, but which is nevertheless able to pick up heat with dramatic effects on the order parameter.
\end{abstract}

\date{\currenttime, \today}

\maketitle

For ultra-cold atoms kept in a magneto-optical trap, Bose-Einstein condensation is a phase transition entirely driven by thermal equilibration \cite{miesner_1998,ritter_2007}. In other bosonic many-body systems this conclusion is sometime harder to reach and requires a careful examination. This is for example the case of photons stored in a cavity filled with dye molecules, for which condensation has been demonstrated \cite{klaers_2010a}. In spite of the intrinsic driven-dissipative (DD) character of the system, it has been found eventually that thermalization happens at a much faster rate than the photon loss rate \cite{klaers_2010b,klaers_2014}, via grand-canonical interaction with the molecules rovibronic degrees of freedom \cite{schmitt_2014}.

Exciton-polaritons (polaritons) in semiconductor microcavities \cite{weisbuch_1992} constitute another example of a nontrivial open system, in which condensation has been reported \cite{kasprzak_2006,balili_2007} and studied. In this case, the DD dynamics usually has a strong influence on the phenomenon \cite{porras_2002,doan_2005,krizha_2007,kasprzak_2008,levrat_2010,snoke_2017}, which thus differs from its textbook thermal-equilibrium counterpart in many respects: condensation can for example take place in more than one state \cite{krizha_2009}, and have a non-zero momentum as a result of broken time reversal symmetry \cite{richard_2005,wouters_2008}, and a diffusive Goldstone mode is expected as the long wavelength excitations \cite{wouters_2007,alessio_2014}. More fundamentally, it has been recently pointed out in a series of theoretical work that owing to its DD character, polariton condensation actually belongs to a universality class which is different from that of equilibrium systems \cite{sieberer_2013}, and which is common to phenomena as diverse as selection in predator-prey dynamics or the build up of traffic jam \cite{diehl_2015,knebel_2015}.

\begin{figure}[t]
\includegraphics[width=0.7\columnwidth]{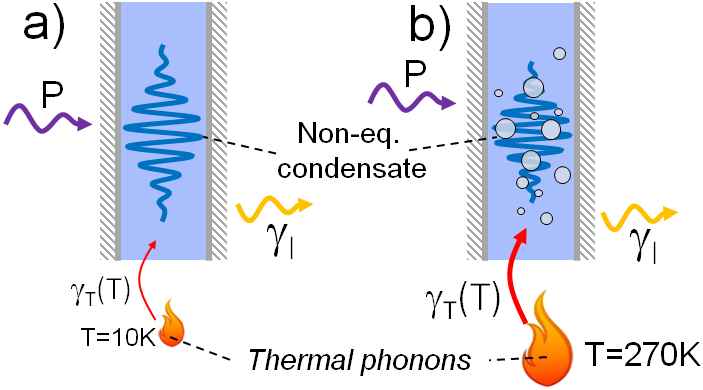}
\caption{(Color online) Schematic representation of two different regimes of polariton condensation. P is the nonresonant pump rate, $\gamma_\mathrm{T}(T)$ is the polariton-thermal phonons inelastic scattering rate, and $\gamma_\ell$ is the overall polariton loss rate. In a) $\gamma_\mathrm{T}\ll\gamma_\ell$ such that condensation can be considered as a dominantly driven-dissipative (DD) phenomena. In b) $\gamma_\mathrm{T}\simeq\gamma_\ell$ such that condensation takes place ``halfway'' between the thermal equilibrium and the DD regimes.}
\label{fig1}
\end{figure}

In this work, we use a warm thermal bath of phonons ($T=150-250\,$K) as a strong heat source interacting with polaritons, to realize and characterize a condensation regime in which, as illustrated in Fig.\ref{fig1}.b, thermal equilibration and the DD dynamics play an equally significant role in the phenomenon, and as a result lies ``halfway'' between both regimes.

%

From a practical point-of-view, this regime is reached when $\gamma_\mathrm{T}$, the polariton scattering rate with thermal phonons, gets comparable with the one-body loss rate $\gamma_\ell$. An important first step is thus to obtain a quantitative measurement of $\gamma_\mathrm{T}$. To do that, we take advantage of the fact that this information is unambiguously encoded within the polariton emission spectral linewidth in the low density, uncondensed regime. With this calibration at hand, we can then turn to the condensate regime and examine how the spatial correlations are affected by the thermal excitations. Another nontrivial aspect of the problem is to find a system in which the polariton fluid can withstand the required warm temperatures. For this purpose, we have fabricated microcavities made up of Zinc Selenide compounds \cite{klembt_2015} in which, like in nitride \cite{levrat_2010}, zinc oxide \cite{li_2013,trichet_2013} or organic material-based microcavities \cite{plumhof_2014}, this requirement is met. In addition, the disorder is weak enough in this system \cite{SI_F} to allow the formation of spatially extended condensates.

\begin{figure}[t]
\includegraphics[width=\columnwidth]{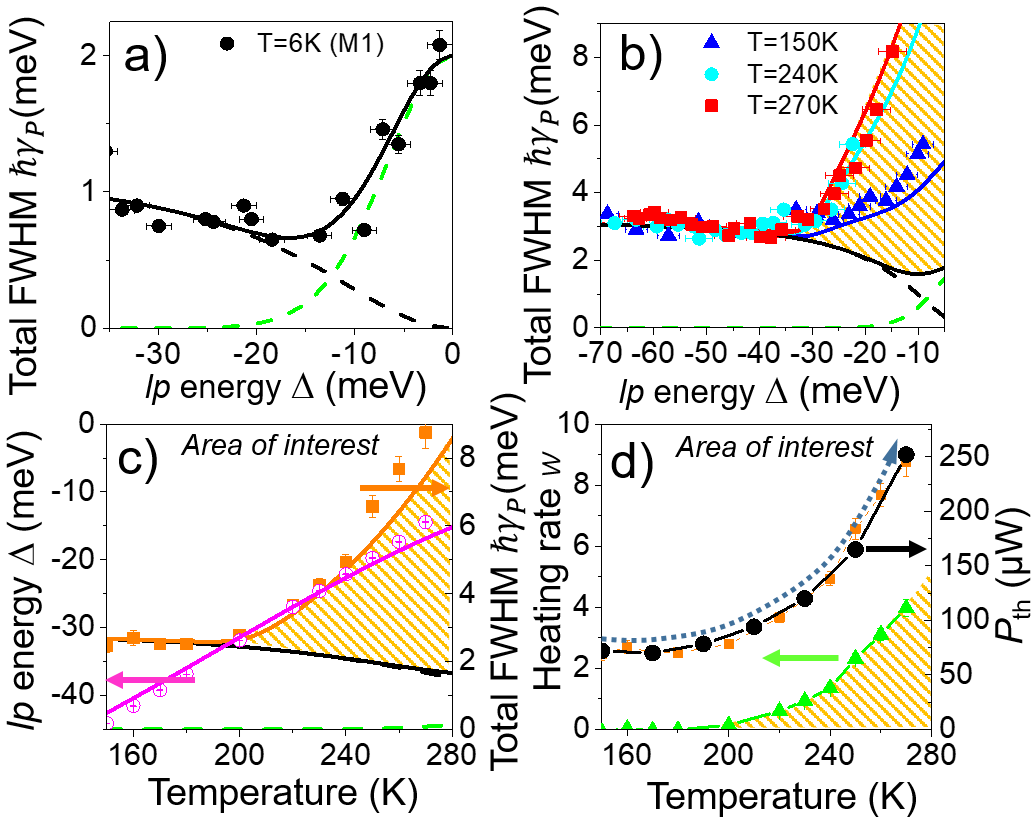}
\caption{(Color online) Measured polariton spectral full-width at half-maximum (FWHM) $\hbar\gamma_\mathrm{p}$ (colored filled circles) versus polariton energy $\Delta=E_\mathrm{lp}-E_\mathrm{hh}$ at a) $T=5\,$K (M1), and b) $T=150\,$K, $240\,$K, and $270\,$K (M2). In a), b), and c), the solid black line, dashed black and dashed green line show the calculated overall polariton loss contribution $\hbar\gamma_\ell =\hbar\gamma_\mathrm{rad}+\hbar\gamma_\mathrm{inh}$ to the FWHM, and both separate contributions respectively. c) Measured $\hbar\gamma_\mathrm{p}$ and $\Delta$ versus temperature at the AOI. d) Measured laser power at threshold $P_\mathrm{th}$ (circles), fit of the latter $\propto \hbar\gamma_\mathrm{p}$ (squares), and normalized heating rate $w$ (triangles) versus temperature at the AOI. The dotted curved arrow shows the trajectory followed in the series of measurements vs temperature shown in Fig.\ref{fig4}. In each panels, the cross-hatched areas show the thermal phonons contribution.}
\label{fig2}
\end{figure}

\begin{figure*}[t]
\centering
\includegraphics[width=\textwidth]{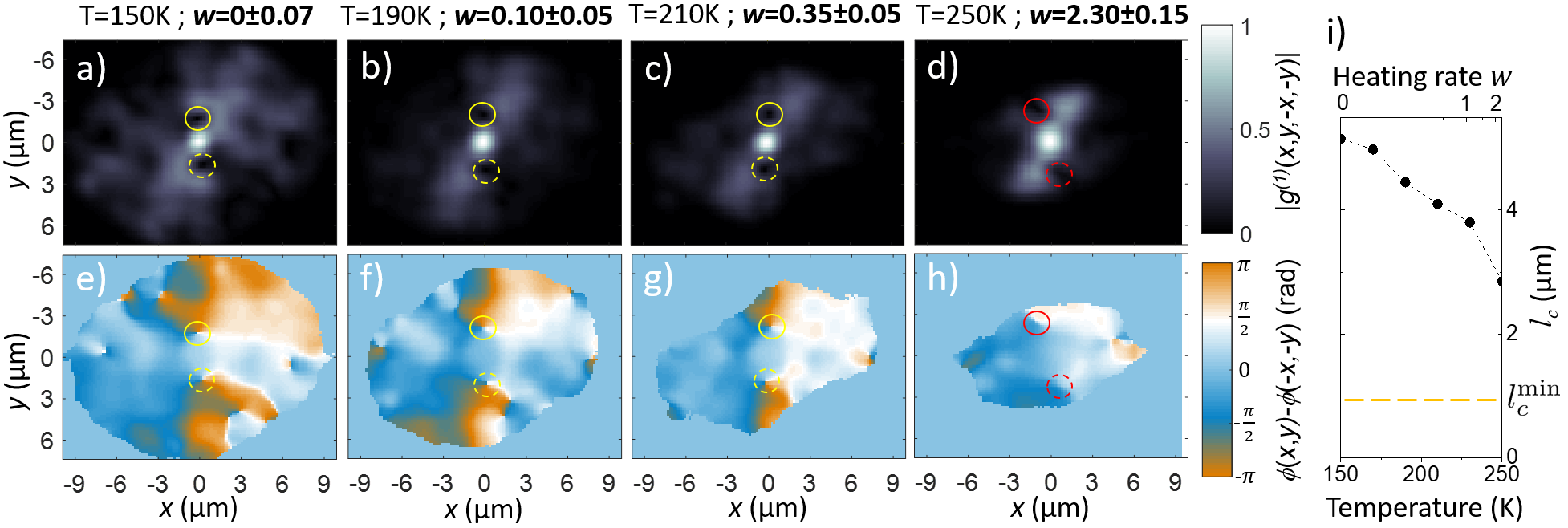}
\caption{(Color online) Polariton condensate correlation amplitude $|g^{(1)}(x,-x,y,-y)|$ (top row), and relative phase $\phi(x,y)-\phi(-x,-y)$ (bottom row) in color scale, versus position $(x,y)$ at $T= 150\,$K (a,e); $190\,$K (b,f); $210\,$K (c,g) and $250\,$K (d,h). The corresponding heating rates $w(T)$ are shown above each column. The yellow (red) circle marks the actual (former) position of a singly-charged charge vortex. i) Measured correlation length vs temperature extracted from measurements a-d. $l_c^\text{min}=$\SI{1}{\micro\meter} is the instrumental lower limit of $l_c$.}
\label{fig3}
\end{figure*}

In order to carry out this experiment, we need two such microcavities M1 and M2, of nominally identical design, but tuned to work in a different temperature range: M1 (M2) has been tuned to work at $T=5K$ ($T=[150,250]\,$K. In both microcavities, a Rabi splitting of $\hbar\Omega_\mathrm{hh}=(32\pm2)\,$meV between the cavity mode and the heavy-hole exciton is achieved. The QW excitonic transition energy decreases from $E_\mathrm{hh}= 2819\,$meV at $T= 10\,$K, to $E_\mathrm{hh}= 2710\,$meV at $T= 270\,$K. Polaritons are excited by non-resonant ($\sim 100\,$meV above $E_\mathrm{hh}$) optical pulses of $1\,$ps duration, generated by a frequency-doubled Ti-Sapphire laser. While this pulsed excitation method does not result in a steady state condensate, it is still well suited to measure the effect of a heat exchange mechanism between polaritons and thermal phonons as its time scale is comparable or even shorter than the polariton lifetime. The experiment is realized in two steps: first, we use both M1 and M2 to extract the different contribution to the polariton emission spectral linewidth in the weak excitation regime, and hence $w=\gamma_\mathrm{T}/\gamma_\ell$, the heating rate to loss rate ratio referred to as the normalized polariton heating rate, versus temperature. Then, we use M2 to measure the complex valued first-order spatial correlation map $g^{(1)}(x,-x,y,-y)$ throughout the crossover between $w\simeq 0$ and $w\geq 1$, which takes place between $T=150\,$K and $250\,$K.



The ground-state polariton emission spectral linewidth $\hbar\gamma_\mathrm{p}$ measurement is carried out under weak excitation, i.e. far below the condensation threshold $P_\mathrm{th}$. $\hbar\gamma_\mathrm{p}$ consists in the sum of three contributions: the polariton-phonon interaction $\hbar\gamma_\mathrm{T}$, the polariton radiative loss $\hbar\gamma_\mathrm{rad}$ that depends on the polaritons photonic fraction, and the quantum well disorder that irreversibly scatter polaritons into localized dark exciton states at a rate $\hbar\gamma_\mathrm{inh}$ \cite{savona_2007}. In order to determine the last two contributions that constitute the overall polariton losses $\hbar\gamma_\ell$, we perform this measurement with M1 at $T=5\,$K where $\hbar\gamma_\mathrm{T}$ vanishes. The measurement is shown in Fig.\ref{fig2}.a versus the polariton energy $\Delta=E_\mathrm{lp}-E_\mathrm{hh}$, where $E_\mathrm{lp}$ is experimentally tunable thanks to the intentional wedged shape of the microcavity spacer.

Both contributions, $\gamma_\mathrm{rad}(\Delta)$ (black dashed line) and $\gamma_\mathrm{inh}(\Delta)$ (green dashed line) to $\gamma_\mathrm{p}$ (solid black line), are well understood within a simple model based on coupled harmonic oscillators to describe the radiative loss rate $\gamma_\mathrm{rad}$, yielding a pure cavity linewidth of $\hbar\gamma_c= 1.2\,$meV, and a Fermi golden rule for the scattering rate by disorder $\gamma_\mathrm{inh}$. For the latter, a Gaussian density of dark states is assumed, with a fitted spectral width of $\hbar\Gamma_\mathrm{inh}= 11\,$meV, and a characteristic scattering rate $R_0= 2\,$meV$/\hbar$ (green dashed line in Fig.\ref{fig2}.a) \cite{SI_A}. Importantly, the three parameters $\Gamma_\mathrm{inh}$, $R_0$ and $\gamma_c$ are temperature independent, and the last two are structural characteristics of the quantum wells and can thus be used also for M2. Due to fabrication uncertainties, $\gamma_c$ is different for M1 and M2, but easily determined by looking at the linewidth of low $\Delta$ polaritons (i.e. of dominantly photonic fraction), for which it is the dominant contribution.

We then measure $\hbar\gamma_\mathrm{p}(\Delta)$ at elevated temperature in M2. The result is shown in Fig.\ref{fig2}.b for three different temperatures, and compared with $\hbar\gamma_\ell(\Delta)$ derived from the low temperature analysis. The difference constitutes the thermal contribution $\hbar\gamma_\mathrm{T}(\Delta,T)$ to $\gamma_\mathrm{p}$, shown as a cross-hatched region. We checked this interpretation with a numerical calculation of the inelastic scattering rate $\gamma_\mathrm{T}(T,\Delta)$ of $k_\parallel=0$ polaritons by thermal phonons using only material parameters known from the literature \cite{SI_B}. As already pointed out \cite{tassone_1997,cassabois_2000}, we find that the contribution of acoustic phonons to this mechanism is negligible as compared to that of longitudinal optical (LO) phonons. The result is added to $\gamma_\ell$ in order to obtain a theoretical $\gamma_\mathrm{p}(\Delta,T)$. As shown in Fig.\ref{fig2}.b, a quantitative agreement with the measurement is reached for all three temperatures.

The goodness-of-fit actually provides us with an upper bound for other possible temperature-dependent contributions than $\gamma_T$ to the polariton linewidth, and hence other sources of perturbation of the condensate coherence, like a temperature-dependent nonradiative relaxation channel for the condensate. While such a contribution cannot be entirely ruled out, the goodness-of-fit shows that it is negligible as compared to $\gamma_T$ as soon as $w\gtrsim 0.5$ (i.e. $T\gtrsim 220\,$K). If not, the calculated $\gamma_\mathrm{p}(T)$ would depart visibly from the measurement in this range. We are thus confident that we have a reliable measurement of $\gamma_\mathrm{T}(T,\Delta)$ and $w(T)$ at hand, and that it is the dominant source of perturbation of the polaritonic field.

In order to measure the effect of heat alone on the first-order spatial correlations $g^{(1)}(x,y,-x,-y)$, we have to work on a fixed $20\times 20$\SI{}{\micro\meter^{2}} area of interest (AOI) of M2 in order to keep the in-plane disordered potential pattern $U_r(x,y)$ fixed for all investigated temperatures. Indeed, changing the position on M2 to compensate for the excitonic energy redshift with the temperature is not an option, as it would result in the sum of two inextricable contributions: the disorder pattern change plus the thermal perturbation of the correlations. We thus chose a fixed AOI, for which a large range of heating rate $w(T)$ can be accessed between $T=150\,$K and $T=270\,$K. The measured emission linewidth $\hbar\gamma_\mathrm{p}(T)$ and polaritons energy $\Delta(T)$ at AOI versus temperature are shown in Fig.\ref{fig2}.c. Both quantities are well explained by our polariton linewidth model with the parameters determined above. $\gamma_\mathrm{T}$ and $\gamma_\ell$ can thus be accurately determined, as well as the resulting $w(T)$ which is plotted in Fig.\ref{fig2}.d, and found to increase from $0$ to $4$ between $T=150\,$K and $T=270\,$K. Interestingly, we see that $w(T)$ is highly nonlinear. This is due to the cumulative effect of both the phonon population growth and the excitonic fraction increase for increasing temperature, plus the large fixed energy of LO phonons \cite{SI_F}.

For the correlation measurement on AOI, an excitation power $P(T)=1.02 P_\mathrm{th}(T)$ is kept throughout all investigated temperatures, so that the condensate fraction fixed by the DD mechanism would remain nominally identical in absence of the heating mechanism. We also checked that the condensate always remains in the single mode regime \cite{krizha_2009}. The measured $P_\mathrm{th}(T)$ is shown in Fig.\ref{fig2}.d, together with the trajectory $P(T)$ represented as a dotted curved arrow. Upon increasing the temperature from $T=150\,$K to $T=270\,$K, $P_\mathrm{th}(T)$ is found to increase by a factor of $\sim 3$, like $\gamma_p(T)$. In order to better understand this feature, we compared the measurement with a rate equation model of polariton lasing detailed in \cite{SI_C} and often used in the literature, in which we assume that only the polariton linewidth depends on the temperature via $\gamma_\mathrm{T}$. In this approach $P_\mathrm{th}(T)\propto\gamma_\mathrm{p}(T)$. As is shown in Fig.\ref{fig2}.d, this relation is very well checked by the data points (overlap of plots with squares and circles symbols), which is a solid confirmation that other parameters such as the excitonic reservoir to polariton condensate relaxation rate, or the excitonic reservoir decay rate \cite{SI_E}, do not vary much as compared to $\gamma_\mathrm{T}(T)$ in this temperature range. Owing to the microscopic meaning of $\gamma_\mathrm{T}$, this threshold increase can thus be traced back to a nonequilibrium condensate depletion mechanism caused by the thermal fluctuations, and reminiscent from the equilibrium phenomenon.

The complex valued map $g^{(1)}(x,y,-x,y)$ is then measured versus $w(T)$, along the trajectory $P(T)$, using an imaging Michelson interferometer arranged in cross-correlation configuration \cite{kasprzak_2006}. The amplitude $|g^{(1)}(x,y,-x,-y)|$, and relative phase $\phi(x,y)-\phi(-x,-y)=\mathrm{Arg}\left\{g^{(1)}(x,y,-x,-y)\right\}$ of the correlation function are extracted and shown in Figs.\ref{fig3}.a-d, and Figs.\ref{fig3}.e-h respectively, for four different temperatures. Note that the data have been clipped at long distances, where the interferogram amplitude has a too low signal-to-noise ratio.

For vanishing $w$ (Fig.\ref{fig3}a-b), the condensate exhibits spatial coherence over a large area, with some amplitude modulation due to the local disorder $U_r(x,y)$ \cite{kasprzak_2006,baas_2008}. For increasing $w$, the long-distance part of the correlations decreases and eventually vanish (Fig.\ref{fig3}c-d). In order to estimate quantitatively this effect, we determined the average correlations length $l_c(T)$ from each maps \cite{SI_D} and plotted it in Fig.\ref{fig3}.i: we see that $l_c$ decreases steadily from \SI{5.31}{\micro\meter} to \SI{2.85}{\micro\meter} for $w$ increasing from $0\pm0.07$ to $2.30\pm0.15$. This decoherence results from the coupling between the condensate and the thermal fluctuations it picks up, which are intrinsically uncorrelated in time and space.

\begin{figure}[t]
\centering
\includegraphics[width=\columnwidth]{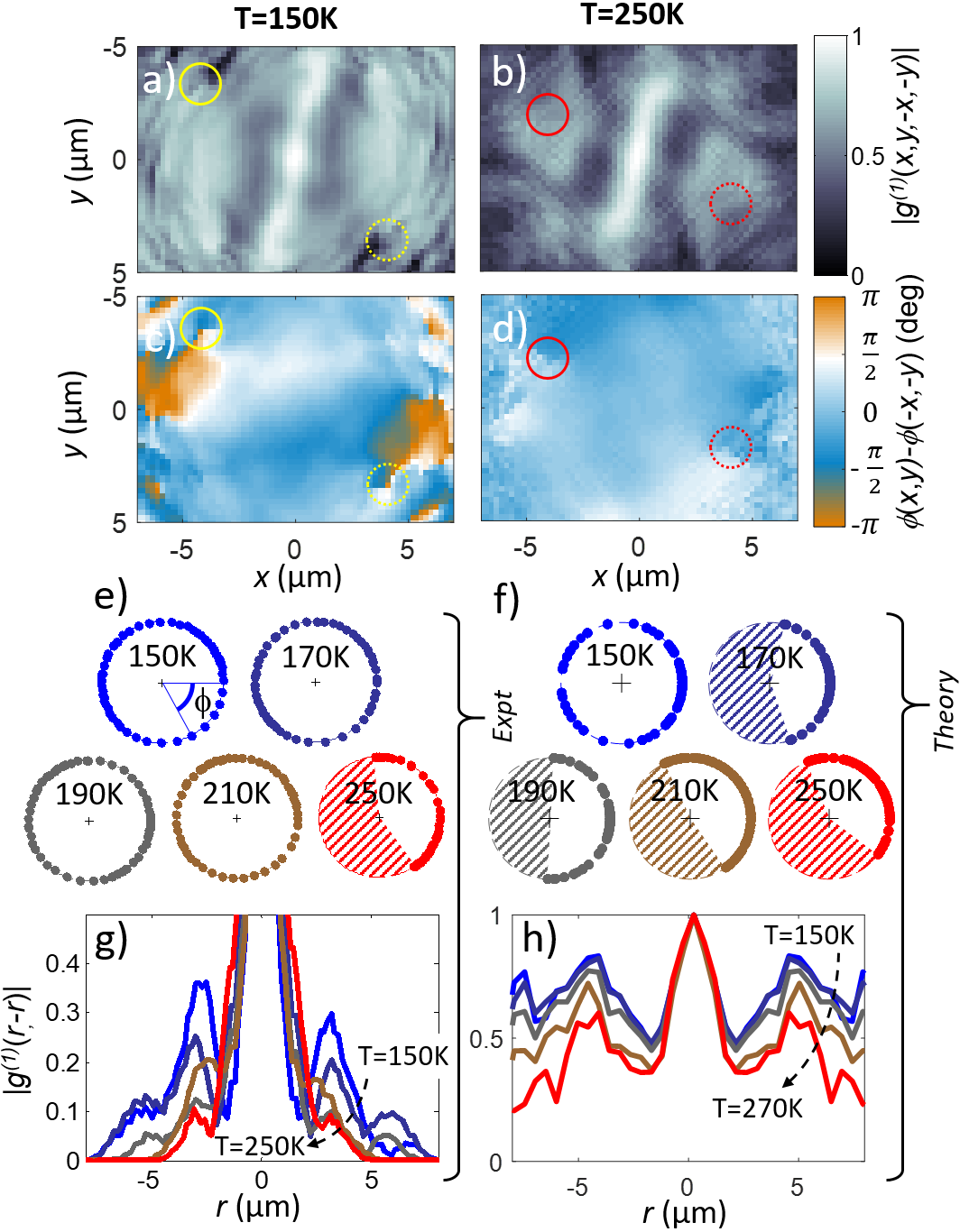}
\caption{(Color online) Calculated Polariton condensate correlation amplitude $|g^{(1)}(x,-x,y,-y)|$ (a-b), and phase $\phi(x,y)-\phi(-x,-y)$ (c-d) versus position $(x,y)$ at $T=150\,$K, and $T=250\,$K. The yellow (red) circles mark the positions of a well-identified (remnants of) vortices. (e) measured and (f) calculated phase values realized around a vortex core at a radius of \SI{1}{\micro\meter}. The cross-hatched region shows the missing phase values due to thermal disconnection. (g) measured and (h) calculated representative cross-sections of $|g^{(1)}(x,-x,y,-y)|$ for $T=\{150,170,190,210,250\}\,$K.}
\label{fig4}
\end{figure}

The measured phase maps shown in Fig.\ref{fig3}.e exhibits a rich flow pattern involving some permanent phase structures, such as a singly-charged quantized vortex and its symmetric counterpart (yellow circles in Fig.\ref{fig3}). To quantitatively follow the stability of these vortices versus the heating rate $w$, we used and measured their defining property: the winding of the condensate phase around their core by an integer times $2\pi$. The results are shown in Fig.\ref{fig4}.e: for $T<250\,$K the condensate phase around the vortex is homogeneously spread within the interval $[0,2\pi]$. At $T=250\,$K ($w=2.30\pm0.15$), we find a large gap of missing values exceeding $\pi$ in this interval, showing that the closed path of the vortex flow is disconnected by the thermal noise. This behaviour is quite the opposite of thermal equilibrium vortices, which exhibit random nucleation and spatial wandering as a result of thermal fluctuations, and thus have a tendency to proliferate upon increasing temperature. Our vortices have a purely DD origin: they are fully deterministic and result from the interplay between DD dynamics and disorder, such that random thermal fluctuations actually disrupts their characteristic flow.

In order to provide a better understanding of these observations, we have simulated the behaviour of the condensed part of the polariton fluid using a time-dependent DD mean field calculation \cite{wouters_2007,carusotto_2013}, including a thermal space- and time-dependent noise source driven by the polariton-phonon interactions as in \cite{savenko_2013}. The model is presented in detail in \cite{SI_E}. We used the same parameters as those used above for the calculation of $\gamma_\mathrm{T}(\Delta,T)$. The equations are solved in time and the obtained first-order correlation map is time-integrated like in the experiment. A realistic polariton disordered potential $U_r(x,y)$ is included, with the amplitude and correlation length obtained from separate measurements.

The calculated correlation amplitude $|g^{(1)}(x,y,-x,-y)|$ are shown in Figs.\ref{fig4}.a-b for $T=150\,$K and $T=250\,$K. A quantitative comparison with the experiment is shown in Fig.\ref{fig4}.g and Fig.\ref{fig4}.h, where measured and calculated cross-sections of $|g^{(1)}(x,y,-x,-y)|$ are plotted for different temperatures. Both show the same trend: a decrease of the correlations at large distance upon increasing temperature, albeit in a less dramatic way in the theory. This discrepancy is expected as, independently from temperature, a mean-field model neglects the non-condensed polaritons population which is significant in our system, and as a result overestimates the long-distance correlation by a constant factor \cite{SI_E}. However, the model is still able to capture qualitatively the main features observed in the experiment, like the shortening of the correlation length and the scrambling of the phase structures when the temperature increases.

The calculated phase maps $\phi(x,y)-\phi(-x,-y)$ are shown in Fig.\ref{fig4}.c-d for $T=150\,$K and $T=250\,$K. A vortex pair is identified and analyzed like in the experiment, and the results are shown in Fig.\ref{fig4}.f. Like in the experiment (Fig.\ref{fig4}.e), a $2\pi$ phase winding disconnection is observed in the calculation above a certain critical temperature as a result of the thermal scrambling of the condensate wavefunction. We investigated further this effect using a few numerically obtained vortices pinned on different disorder patterns \cite{SI_G}. We found that the disconnection occurs at about the same temperature for all of them, but the size of the phase gap seems to depend on the details of the disorder in a nontrivial way.

In summary, using hot thermal phonons as a controlled and measurable source of heat, we have achieved and characterized a regime of polariton condensation situated ``halfway'' between the thermal equilibrium and the driven-dissipative regimes. These properties have reminiscence from both regimes: upon increasing $w$, a condensate thermal depletion channel grows in contribution, the first-order correlation length shrinks, and the driven-dissipative vortices get disconnected by the thermal fluctuations. How much heat is actually required to scramble such structures, depending on their specific topology and on the local disorder, is an important perspective considering realistic devices based on topological protection \cite{karzig_2015}. More fundamentally, our observation opens up profound questions in the context of non-equilibrium phase transitions, e.g. what is the identity and characteristics of the universality class describing condensation in a system subjected to both DD dynamics and thermal fluctuations, with a continuously variable relative contributions $w$.

\begin{acknowledgments}
The authors acknowledge financial support from the ERC (No. 258608), and the french ANR (No. ANR-16-CE30-0021) and LANEF framework (No. ANR-10-LABX-51-01). We thank D. Hommel and C. Kruse for the microcavity fabrication. Enlightening discussions with A. Auff\`eves, I. Carusotto, A. Chiocchetta, C. Elouard, M. Wouters, and D. Squizzato are acknowledged. MR wishes to express his immense gratitude to B. Richard for her invaluable support. The first two authors contributed equally to this work.
\end{acknowledgments}

\end{document}